\documentclass[prd,onecolumn,amsmath,amssymb,floatfix,nofootinbib]{revtex4}

\usepackage{amssymb}
\usepackage{amsmath}
\usepackage{slashed}

\usepackage{caption}
\usepackage{subfigure}

\usepackage{graphicx}
\usepackage{graphics}
\usepackage{dcolumn}
\usepackage{color}
\usepackage{rotate}
\usepackage{fancyhdr}
\usepackage{hyperref}
\hypersetup{
    colorlinks=true,
    linkcolor=black,
    filecolor=black,
    urlcolor=black,
}
\usepackage{indentfirst}
\usepackage[titletoc,title]{appendix}

\def\be{\begin{eqnarray}}
\def\ee{\end{eqnarray}}
\def\no{\nonumber}

\def\VEV#1{\left\langle #1 \right\rangle}

\definecolor{darkred}{rgb}{.743,0,0}

\def\lsim{\mathrel{\rlap{\lower4pt\hbox{\hskip1pt$\sim$}}
     \raise1pt\hbox{$<$}}}         
\def\gsim{\mathrel{\rlap{\lower4pt\hbox{\hskip1pt$\sim$}}
     \raise1pt\hbox{$>$}}}         
\newcommand{\beq}{\begin{equation}}
\newcommand{\eeq}{\end{equation}}

\newenvironment{Eqnarray}{\arraycolsep 0.14em\begin{eqnarray}}{\end{eqnarray}}
\def\beqa{\begin{Eqnarray}}
\def\eeqa{\end{Eqnarray}}

\begin{document}
\title{Large Higgs-electron Yukawa coupling in 2HDM}

\author{Avital Dery$^{1a}$, Claudia Frugiuele$^{1a}$ and Yosef Nir$^{1a}$}
\affiliation{$^1$Department of Particle Physics and Astrophysics, Weizmann Institute of Science, Rehovot, Israel 7610001}
\email{$^a$avital.dery, claudia.frugiuele, yosef.nir@weizmann.ac.il}

\begin{abstract}
\noindent
The present upper bound on $\kappa_e$, the ratio between the electron Yukawa coupling and its Standard Model value, is of ${\cal O}(600)$. We ask what would be the implications in case that $\kappa_e$ is close to this upper bound. The simplest extension that allows for such enhancement is that of two Higgs doublet models (2HDM) without natural flavor conservation. In this framework, we find the following consequences: (i) Under certain conditions, measuring $\kappa_e$ and $\kappa_V$ would be enough to predict values of Yukawa couplings for other fermions and for the $H$ and $A$ scalars. (ii) In the case that the scalar potential has a softly broken $Z_2$ symmetry, the second Higgs doublet must be light, but if there is hard breaking of the symmetry, the second Higgs doublet can be much heavier than the electroweak scale and still allow the electron Yukawa coupling to be very different from its SM value. (iii) CP must not be violated at a level higher than ${\cal O}(0.01/\kappa_e)$ in both the scalar potential and the Yukawa sector. (iv) LHC searches for $e^+e^-$ resonances constrain this scenario in a significant way. Finally, we study the implications for models where one of the scalar doublets couples only to the first generation, or only to the third generation.
\end{abstract}

\maketitle


\vskip 8pt

%
%

\vglue 0.3truecm


\section{Introduction}
The Standard Model (SM) predicts the values of the Yukawa couplings. The diagonal couplings are proportional to the corresponding fermion masses,
\be
y_f\equiv Y_{ff}^{\rm SM}=\sqrt{2} m_f/v,
\ee
while off-diagonal couplings vanish. As concerns the diagonal couplings, the LHC experiments are testing the SM predictions $\kappa_f=1$ and $\tilde\kappa_f=0$, where
\be
\kappa_f\equiv& {\cal R}e(Y_{ff}/Y_{ff}^{\rm SM}),\no\\
\tilde\kappa_f\equiv& {\cal I}m(Y_{ff}/Y_{ff}^{\rm SM}).
\ee
While measurements of the third generation Yukawa couplings imply that $\kappa_{t,b,\tau}={\cal O}(1)$ \cite{Aaboud:2017xsd,Sirunyan:2017khh,Aad:2015vsa,Khachatryan:2016vau}, direct measurements still allow the Yukawa couplings of the first two generations to be very different from the SM values. For the second generation, there is a mild upper bound on the charm Yukawa \cite{Perez:2015aoa}, $\kappa_c\lsim 17$, and a significant constraint on the muon Yukawa \cite{Aaboud:2017ojs}, $\kappa_\mu\lsim1.7$. The first generation Yukawa couplings can still be orders of magnitude larger than their SM values. Moreover, there is theoretical motivation to consider a different source for the Yukawa couplings of the first two (or just the first) generations that would explain their smallness (see. {\it e.g.}, \cite{Botella:2016krk,Ghosh:2015gpa,Altmannshofer:2015esa}). We ask here whether indeed the Yukawa couplings of light fermions could be very different from their SM values, {\it i.e.} $\kappa\gg1$ or $\kappa\ll1$. For concreteness, we study the lightest charged fermion, $\kappa_e$.

As concerns the Yukawa coupling of the electron, the two most constraining measurements are the CMS bound on $h\to e^+e^-$ \cite{Khachatryan:2014aep},
\be\label{eq:muee}
\mu_{ee}\equiv\frac{\sigma(pp\to h){\rm BR}(h\to e^+e^-)}{[\sigma(pp\to h){\rm BR}(h\to e^+e^-)]^{\rm SM}}<3.7\times10^5,
\ee
and the ACME bound on the electron EDM \cite{Baron:2013eja},
\be\label{eq:deexp}
|d_e|<8.7\times10^{-29}\ e\ {\rm cm}.
\ee
With $\kappa_t\sim1$ \cite{Barr:1990vd}, these bounds translate into \cite{Brod:2013cka,Altmannshofer:2015qra}
\be\label{eq:keexp}
|\kappa_e|\leq 6.1\times10^2,\ \ \ |\tilde\kappa_e| \leq 1.7  \times 10^{-2}.
\ee
For the sake of concreteness, we consider the hypothetical case that $\kappa_e$ is close to the experimental bound, $\kappa_e={\cal O}(500)$.

A significant deviation of $\kappa_e$ from unity is most simply accounted for in models  with more than one Higgs doublet. Hence, we consider two Higgs doublet models (2HDM). For a review of this framework, see Ref.~\cite{Branco:2011iw}. 2HDM with natural flavor conservation (NFC) predict $\kappa_e=\kappa_\mu=\kappa_\tau$. The measurement of $\mu_{\tau\tau}$ \cite{Sirunyan:2017khh,Aad:2015vsa},
\be
\mu_{\tau\tau}=1.09\pm0.23,
\ee
as well as the upper bound on $\mu_{\mu\mu}$ \cite{Aaboud:2017ojs},
\be\label{eq:mumumu}
\mu_{\mu\mu}<2.8,
\ee
thus exclude the possibility that $\kappa_e\gg1$ for NFC models. Hence, we consider 2HDM without NFC.

The plan of this paper is as follows. In Section \ref{sec:yukawa} we define a basis for the two scalar doublets that is particularly convenient for our purposes. We find conditions under which the Yukawa couplings of fermions to the light scalar $h$, and to the heavy scalars $A$, $H$ and $H^\pm$, are related to that of the electron. In Section \ref{sec:spectrum} we study the implications of a very large or very small $\kappa_e$ for the scalar spectrum. In Section \ref{sec:cpv} we obtain constraints on CP violation in the scalar potential when $\kappa_e$ is enhanced. We further re-analyze the one-loop contributions to $d_e$ for large Yukawa coupling. Section \ref{sec:lhc} is devoted to discussion of the LHC phenomenology. In Section \ref{sec:models} we survey two Higgs doublet models in the literature for which our results provide further insights. We conclude in Section \ref{sec:conclusions}.

\section{Yukawa couplings in 2HDM}
\label{sec:yukawa}

\subsection{$\kappa_e$ in the ``$\beta_e$-basis"}\label{sec:be}
In this section, we assume that CP is a good symmetry of the scalar potential and of the Yukawa sector. Later we argue that this is actually a requirement (rather than an assumption) if $\kappa_e\gg1$. We use notations and various relations based on Ref. \cite{Dery:2013aba}.

In the Higgs basis, $(\Phi_M,\Phi_A)$, defined by
\be
\VEV{\Phi_M}=v,\ \ \ \VEV{\Phi_A}=0,
\ee
and in the mass basis for the charged leptons,  we have
\be
Y^M={\rm diag}(y_e,y_\mu,y_\tau),\ \ \
Y^A={\rm arbitrary}.
\ee
(Since we deal mostly with the charged lepton sector, we use the notation $Y^X$ for the charged lepton Yukawa matrix of $\Phi_X$.)

The Yukawa matrices of the neutral CP-even scalars are given by
\be\label{eq:higgstomass}
Y^h=&c_{\alpha-\beta}Y^A+s_{\beta-\alpha}Y^M,\no\\
Y^H=&s_{\alpha-\beta}Y^A+c_{\beta-\alpha}Y^M,
\ee
where $\alpha-\beta$ is the rotation angle from the $(\Phi_M,\Phi_A)$ basis to the $(\Phi_H,\Phi_h)$ basis. Here, and in what follows, we use $c_\phi,s_\phi,t_\phi,\cot_\phi$ for, respectively, $\sin\phi,\cos\phi,\tan\phi,\cot\phi$. Defining $y_A^e\equiv Y^A_{ee}$, we obtain
\be
\kappa_e=s_{\beta-\alpha}+c_{\beta-\alpha}(y_A^e/y_e).
\ee

Thus
\begin{align}
|\kappa_e|\gg1&\ \Longrightarrow\ |c_{\beta-\alpha}(y_A^e/y_e)|\gg1,\nonumber\\
|\kappa_e|\ll1&\ \Longrightarrow\ |1+\cot_{\beta-\alpha}(y_A^e/y_e)|\ll1.
\end{align}

We now rotate to a basis for the scalar doublets, $(\Phi_1,\Phi_2)$, that is rotated by an angle $\beta$ from the Higgs basis. We define $y_1^e=Y^1_{ee}$ and $y_2^e=Y^2_{ee}$. We obtain:
\be
\frac{y^A_e}{y_e}=\frac{-s_\beta y_1^e+c_\beta y_2^e}{c_\beta y_1^e+s_\beta y_2^e}.
\ee
Things are simplified in a specific basis, where $y_2^e=0$. We can always find a rotation angle, $\beta=\beta_e$:
\be
\beta_e=-\arctan(y_A^e/y_e),
\ee
that takes us to this basis. Then,
\be\label{eq:keba}
\kappa_e=s_{\beta-\alpha}-c_{\beta-\alpha}t_{\beta_e}.
\ee

To have $|\kappa_e|\gg1$, we need
\be
|c_{\beta-\alpha}t_{\beta_e}|\gg1.
\ee
To have $|\kappa_e|\ll1$, we need
\be
\tan\alpha_e\ll\cot\beta_e.
\ee
Things are simplified if $|c_{\beta-\alpha}|\ll1$ or, in other words, $\kappa_V\simeq1$. In this case we have the following scenarios:
\begin{enumerate}
\item $|\kappa_e|\gg1$ requires $t_{\beta_e}\gg t_{\beta-\alpha}$: $\beta_e$ closer to $\pi/2$ than $\beta-\alpha$ means that $\Phi_M\simeq\Phi_2$ (the doublet that has no $ee$ coupling).
\item $|\kappa_e|\ll1$ requires $t_{\beta_e}\simeq t_{\beta-\alpha}$: $\beta_e$ close to $\beta-\alpha$ means that $\Phi_h\simeq\Phi_2$.
\item $|\kappa_e|\simeq1$ requires $t_{\beta_e}\ll t_{\beta-\alpha}$ or $t_{\beta_e}\simeq 2t_{\beta-\alpha}$: $\beta_e$ close to $0$ means that $\Phi_M\simeq\Phi_1$.
\end{enumerate}

\begin{figure}[t]
 \begin{center}
  \includegraphics[width=0.90\textwidth]{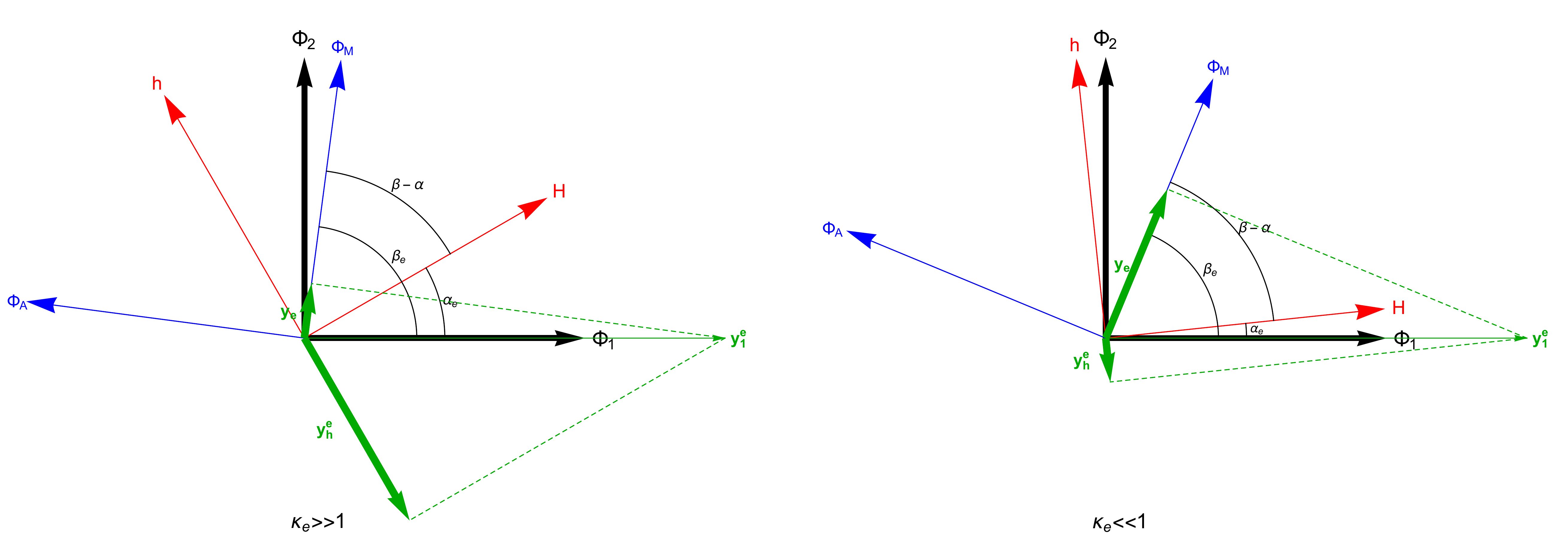} \
 \end{center}
     \caption{Geometric representation of the three bases, $(\Phi_1,\Phi_2)$ (black), $(\Phi_M,\Phi_A)$ (blue), and $(\Phi_h,\Phi_H)$ (red).
 {\bf Left:} $\kappa_e\gg1$. {\bf Right:} $\kappa_e\ll1$.}
\label{fig:bases}
\end{figure}

The rotations between the three bases that we use for the two Higgs doublets are presented in Fig. \ref{fig:bases}.
%

\subsection{$\kappa_f$ for $f\neq e$}
\label{sec:kappaf}
In this subsection we identify conditions under which the diagonal couplings of fermions $f\neq e$ are related to that of the electron. For this purpose, we employ the $\beta_e$-basis, defined in the previous section. In general, this basis plays no special role for the other fermions, and both $y_1^f$ and $y_2^f$ are different from zero. In such a case, there is no further predictive power for $\kappa_f$ (unless a specific flavor model is assumed). There are, however, two special cases in which there is a strong predictive power:
\begin{enumerate}
\item  Similarly to the electron, $y_2^f=0$. In this case,
\be\label{eq:classone}
\kappa_f=\kappa_e.
\ee
\item In contrast to the electron, $y_1^f=0$. In this case,
\be\label{eq:classtwo}
\kappa_V=\frac{1+\kappa_e\kappa_f}{\kappa_e+\kappa_f}.
\ee
In particular, for $\kappa_e$ very different from 1,
\be
k_f\simeq\left\{\begin{matrix} \kappa_V & \kappa_e\gg1\\
\kappa_V^{-1} & \kappa_e\ll1.\end{matrix}\right.
\ee
Thus, if $y_2^e=0$ and $y_1^f=0$, then $\kappa_e\gg1$ or $\ll1$ implies $\kappa_f\approx1$.
\end{enumerate}
Both of these classes are demonstrated by NFC types II,III,IV. The first case is demonstrated by NFC type I. (See Ref. \cite{Branco:2011iw} for a review of the various NFC models.) Our findings here are, however, much more general than NFC. They apply whenever, in the fermion mass basis and some basis for the two Higgs doublets, two (or more) diagonal entries vanish. In particular, Eqs. (\ref{eq:classone}) and (\ref{eq:classtwo}) are independent on whether off-diagonal terms and other diagonal terms vanish (as they do in NFC models) or not.

\subsection{$\kappa_f^X$ for $X=A,H$}
In this subsection we identify conditions under which the diagonal Yukawa couplings of the $H$ and $A$ are related to $Y^h_{ee}$. We define
\be
\kappa_f^{A,H}={\cal R}e(Y^{A,H}_{ff}/y_f),
\ee
For the pseudoscalar $A$, we use the relations between the $\beta_e$-basis and the Higgs basis:
\be
Y^M=&+c_{\beta_e} Y^1+s_{\beta_e} Y^2,\no\\
Y^A=&-s_{\beta_e} Y^1+c_{\beta_e} Y^2.
\ee
For the scalar $H$, we use the relation between the mass basis and the Higgs basis of Eq. (\ref{eq:higgstomass}), which leads to
\be\label{eq:kHA}
\kappa_f^H=s_{\alpha-\beta}\kappa_f^A+c_{\alpha-\beta}.
\ee
Again, for fermions with $y_1^f\neq0$ and $y_2^f\neq0$, there is no predictive power for $\kappa_f^{A,H}$, but in the other two cases, there is a strong predictive power:
\begin{enumerate}
\item $y_2^f=0$: Using Eq. (\ref{eq:keba}), we obtain
\be\label{eq:keA}
\kappa_f^A&=&-\tan\beta=\frac{s_{\alpha-\beta}+\kappa_e}{c_{\alpha-\beta}},\no\\
\kappa_f^H&=&t_{\alpha-\beta}(s_{\alpha-\beta}+\kappa_e)+c_{\alpha-\beta}.
\ee
For very large or very small $\kappa_e$, we obtain
\be\label{eq:keAlarge}
k_f^A&\simeq&\left\{\begin{matrix}  \frac{\kappa_e}{c_{\alpha-\beta}}& \kappa_e\gg1\\
t_{\alpha-\beta} & \kappa_e\ll1.\end{matrix}\right.,\no\\
k_f^H&\simeq&\left\{\begin{matrix}  t_{\alpha-\beta}\kappa_e& \kappa_e\gg1\\
t_{\alpha-\beta}s_{\alpha-\beta} & \kappa_e\ll1.\end{matrix}\right.
\ee
Note that these results always apply to $\kappa_e^{A,H}$.
\item $y_1^f=0$: Using Eq. (\ref{eq:keba}), we obtain
\be\label{eq:ktA}
\kappa_f^A&=&\cot\beta=\frac{c_{\beta-\alpha}}{s_{\beta-\alpha}-\kappa_e},\no\\
\kappa_f^H&=&\frac{c_{\alpha-\beta}\kappa_e}{s_{\alpha-\beta}+\kappa_e}.
\ee
For very large or very small $\kappa_e$, we obtain
\be
k_f^A&\simeq&\left\{\begin{matrix}  -\frac{c_{\beta-\alpha}}{\kappa_e}& \kappa_e\gg1\\
t_{\beta-\alpha}^{-1} & \kappa_e\ll1.\end{matrix}\right.,\no\\
k_f^H&\simeq&\left\{\begin{matrix}  c_{\beta-\alpha}& \kappa_e\gg1\\
\cot_{\alpha-\beta}\kappa_e & \kappa_e\ll1.\end{matrix}\right.
\ee
\end{enumerate}
We learn the following lessons:
\begin{itemize}
\item For any fermion with enhanced diagonal Yukawa coupling to the $h$ scalar (such as we assume for the electron in this work), the diagonal Yukawa couplings to the $H$ and $A$ scalars are even further enhanced: $\kappa_f^{A,H}/\kappa_f={\cal O}(t_{\beta-\alpha})$.
\item For any fermion with suppressed diagonal Yukawa coupling to the $h$ scalar, the diagonal Yukawa couplings to the $H$ and $A$ scalars are not suppressed, $\kappa_f^{A,H}={\cal O}(t_{\beta-\alpha})$.
\item If, in some basis for the scalar doublets, $Y^2_{ii}=0$, $Y^1_{jj}=0$, and $\kappa_i\gg1$, then $\kappa_j^A$ is highly suppressed and $\kappa_j^H$ is suppressed but more mildly.
\item If, in some basis for the scalar doublets, $Y^1_{ii}=0$, $Y^2_{jj}=0$, and $\kappa_i\ll1$, then $\kappa_j^H$ is highly suppressed and $\kappa_j^A$ is suppressed but more mildly.
\end{itemize}

\section{The scalar spectrum}
\label{sec:spectrum}
It is  well known that in the decoupling limit of the 2HDM, $m_A^2\gg v^2$, all the light Higgs boson couplings converge to their SM values. Thus, a large deviation of $\kappa_e$ from 1, the case of interest in this work, seems to require that the second Higgs doublet is not very heavy. In this section we investigate whether there are caveats to this statement, and whether special relations within the scalar spectrum are required for strong enhancement of $\kappa_e$. We use the formalism and equations of Ref. \cite{Gunion:2002zf}, but apply them specifically in the $\beta_e$-basis. Thus, in this section, $\beta$ and $\alpha$ stand for $\beta_e$ and $\alpha_e$.

The scalar potential is given by
\begin{align}\label{eq:scapot}
V=&m_1^2(\Phi_1^\dagger\Phi_1)+m_2^2(\Phi_2^\dagger\Phi_2)
-\left[m_{12}^2(\Phi_1^\dagger\Phi_2)+{\rm h.c.}\right]\\
&+\frac{\lambda_1}{2}(\Phi_1^\dagger\Phi_1)^2+\frac{\lambda_2}{2}(\Phi_2^\dagger\Phi_2)^2
+\lambda_3(\phi_1^\dagger\Phi_1)(\Phi_2^\dagger\Phi_2)
+\lambda_4(\phi_1^\dagger\Phi_2)(\phi_2^\dagger\Phi_1)\nonumber\\
&+\left\{\frac{\lambda_5}{2}(\Phi_1^\dagger\Phi_2)^2
+\left[\lambda_6(\Phi_1^\dagger\Phi_1)+\lambda_7(\Phi_2^\dagger\Phi_2)\right]\Phi_1^\dagger\Phi_2
+{\rm h.c.}\right\}\nonumber.
\end{align}
In a general, but CP conserving, 2HDM, the masses-squared of the CP-odd neutral and the charged scalar are given by
\begin{align}
m_A^2&=\frac{m_{12}^2}{s_\beta c_\beta} -\frac12 v^2(2\lambda_5+\lambda_6 t_\beta^{-1} +\lambda_7 t_\beta),\nonumber\\
m_{H^\pm}^2&=m_A^2+\frac12 v^2(\lambda_5-\lambda_4).
\end{align}
The mass-squared matrix for the neutral CP-even Higgs bosons, $H$ and $h$, is given by
\begin{align}
{\cal M}^2&=m_A^2\left(\begin{matrix} s_\beta^2 & -s_\beta c_\beta\\ -s_\beta c_\beta & c_\beta^2\end{matrix}\right)
+{\cal B}^2,\nonumber\\
{\cal B}^2&=v^2\left(\begin{matrix} \lambda_1 c_\beta^2+2\lambda_6 s_\beta c_\beta +\lambda_5 s_\beta^2 &
\lambda_{34}s_\beta c_\beta+\lambda_6 c_\beta^2+\lambda_7 s_\beta^2\\
\lambda_{34}s_\beta c_\beta+\lambda_6 c_\beta^2+\lambda_7 s_\beta^2 &
\lambda_2 s_\beta^2+2\lambda_7 s_\beta c_\beta +\lambda_5 c_\beta^2 \end{matrix}\right),
\end{align}
where $\lambda_{34}\equiv\lambda_3+\lambda_4$.

Eq. (\ref{eq:keba}) implies that $\kappa_e\gg1$ requires $t_{\beta_e}\gg1$, and that for $c_{\alpha-\beta}\ll1$ also $\kappa_e\ll1$ requires $t_{\beta_e}\gg1$. We thus take the $t_\beta\gg1$ and $c_{\beta-\alpha}\ll1$ limits.

\subsection{$m_A^2\gg v^2$}
The main question in our mind is whether measuring $\kappa_e\gg1$ will imply that a second Higgs doublet is necessarily within the reach of the LHC experiments. Thus, we expand the above expressions in the $m_A^2\gg v^2$ limit. We obtain the following relation between $\kappa_e$ and the scalar related parameters:
\be\label{eq:kel7}
\kappa_V-\kappa_e\simeq\frac{\lambda_7 v^2 t_\beta}{m_A^2}.
\ee
We learn that there is an interesting range, $v^2\ll m_A^2\ll v^2 t_\beta$, where $\kappa_e\gg1$ is possible, yet the second Higgs doublet is too heavy to be directly produced by the LHC. It is important to remember that $t_{\beta_e}$ can be very large. Perturbativity requires that $t_{\beta_e}\lsim1/y_e$ (four orders of magnitude above the bound in, for example, NFC type-II models, $t_\beta\lsim1/y_b$). In fact, to achieve $\kappa_e={\cal O}(500)$ with $c_{\alpha-\beta}\lsim0.1$, we need $t_{\beta_e}\gsim5000$.

The larger $m_A^2$, the smaller $c_{\alpha-\beta}\approx \lambda_7 v^2/m_A^2$, which leads to larger $y^A_e$. Requiring that $y^A_e$ is perturbative implies, for large $\kappa_e$, $m_A^2\lsim v^2/(y_e\kappa_e)$. We conclude that, for large $\kappa_e$, $m_A^2$ can be of ${\cal O}[v^2/(y_e\kappa_e)]\gg v^2$. For $\kappa_e={\cal O}(500)$, we can have $m_A={\cal O}(10\ {\rm TeV})$.

It is interesting to understand how this scenario translates into the language of the SM as an effective field theory (EFT). In general, if there are no new light degrees of freedom, modifications of the electron Yukawa coupling come from higher-dimensional operators. Consider the terms
\be\label{eq:nonren}
\lambda_{ij}\overline{L_i}\phi E_j+\frac{\lambda^\prime_{ij}}{\Lambda^2}(\phi^\dagger\phi)\overline{L_i}\phi E_j,
\ee
where $\phi$ is the Higgs doublet, $L_i$ are the left-handed lepton doublets and $E_j$ are the right-handed charged lepton singlets. Then,
\begin{align}
m_e&=\frac{v}{\sqrt2}\left(\lambda_{ee}+\frac{v^2\lambda^\prime_{ee}}{2\Lambda^2}\right),\nonumber\\ Y^h_{ee}&=\lambda_{ee}+\frac{3v^2\lambda^\prime_{ee}}{2\Lambda^2}.
\end{align}
In general, we expect $\kappa_e$ to be between 1, which corresponds to the case that the renormalizable term dominates, and $3$, which corresponds to the case that the dimension-six term dominates. It is possible, however, to have $\kappa_e\gg1$, if $\lambda_{ee}+(\lambda^\prime_{ee}/2)(v^2/\Lambda^2)\ll\lambda_{ee}$. Indeed, this is what is happening in our 2HDM. We obtain, for $m_{11}^2\gg m_{12}^2,m_{22}^2$ and large $\tan\beta_e$,
\begin{align}
\lambda_{ee}&=\sqrt2\left(Y^h_{ee}+\frac{Y^H_{ee}m_{Hh}^2}{m_{HH}^2}\right),\nonumber\\
\frac{v^2\lambda^\prime_{ee}}{2\Lambda^2}&=-\frac{Y^H_{ee}\lambda_{Hhhh} v^2}{\sqrt2 m_{HH}^2}.
\end{align}
Here, $m_{HH}^2$, $m_{Hh}^2$ and $\lambda_{Hhhh}$ are parameters of the scalar potential in the mass basis $(\Phi_H,\Phi_h)$, corresponding to $m_{11}^2$, $m_{12}^2$ and $\lambda_7$, respectively. The mass basis parameters fulfill $m_{Hh}^2\approx(3/2)\lambda_{Hhhh} v^2$. Moreover, in the region of interest, $Y^h_{ee}/Y^H_{ee}=-c_{\beta-\alpha}\approx-\lambda_{Hhhh} v^2/m_{HH}^2$. Consequently, the required cancelation occurs.

\subsection{$\lambda_6=\lambda_7=0$}\label{sec:l6l7}
In various models, zeros in the Yukawa matrices are generated by a $Z_2$ symmetry, under which one of the scalar doublets is even and the other is odd. For phenomenological reasons, the symmetry is usually assumed to be softly broken, namely $m^2_{12}=0$ while $\lambda_6=\lambda_7=0$. For this class of models, We obtain the following relation between $\kappa_e$ and the scalar related parameters:

\be\label{eq:spectrum67}
\frac{m_A^2-\lambda_{34}v^2}{m_H^2-m_h^2}=\frac{\kappa_e(\kappa_V\kappa_e-1)}{\kappa_e-\kappa_V}.
\ee

We learn the following:
\begin{enumerate}
\item $\kappa_e\gg1$ requires
\be\label{eq:keg67}
m_H^2\simeq m_h^2.
\ee
\item $\kappa_e\ll1$ requires
\be\label{eq:kes67}
m_A^2\simeq\lambda_{34}v^2.
\ee
\end{enumerate}
In either case, $m_A^2={\cal O}(v^2)$.

In fact, for $m_A^2\gg v^2$ we have $\kappa_e=1+{\cal O}(v^2/m_A^2)$. To understand why this is the case even for $v^2\ll m_A^2\ll v^2\tan\beta$, we use various equations of Ref. \cite{Gunion:2002zf} and obtain, for $m_A^2\gg v^2$ and $\lambda_6=\lambda_7=0$:
\be
c_{\beta-\alpha}t_\beta=(\lambda_{34}+\lambda_5-\lambda_2)(v^2/m_A^2).
\ee
In other words, in the limit of $m_A^2\gg v^2$ and $\tan\beta\gg1$, $\cos(\beta-\alpha)$ is further suppressed below $(v^2/m_A^2)$, in such a way that $c_{\beta-\alpha}t_\beta$ is small, ${\cal O}(v^2/m_A^2)$, and $\kappa_e$ is consequently ${\cal O}(1)$.

\section{CP violation}
\label{sec:cpv}
Given that we consider the hypothetical case of $\kappa_e \sim 500$, and that there is a bound on $\tilde\kappa_e < 0.017$ \cite{Altmannshofer:2015qra}, CP must be a very good symmetry (broken at a level smaller than $10^{-4}-10^{-5}$) in this context. Here we investigate CP violation in the scalar potential and in the Yukawa couplings, assuming no cancelations between these two sources of $\tilde\kappa_e\neq0$. (For previous, related studies, see Refs. \cite{Inoue:2014nva,Barger:1996jc,Weinberg:1990me,Blankenburg:2012ex,Harnik:2012pb,Broggio:2014mna}.)

\subsection{The scalar sector}
We use here the formalism of Ref. \cite{Inoue:2014nva}. Consider the scalar potential of Eq. (\ref{eq:scapot}) where, for simplicity, we take $\lambda_6=\lambda_7=0$.  We work in the basis where $v_1$ is real. We define
\begin{align}
\tan\beta=&|v_2/v_1|,\\
\mu_{12}=&{\cal R}e(m_{12}^2)/(v^2c_\beta s_\beta),\nonumber\\
\lambda_{345}=&\lambda_3+\lambda_4+{\cal R}e(\lambda_5).\nonumber
\end{align}
The scalar potential gives the following mass-squared matrix in the $\{H_1^0,H_2^0,A^0\}$ basis:
\be
{\cal M}^2=v^2\left(\begin{matrix}
\lambda_1 c_\beta^2+\mu_{12}s_\beta^2 & (\lambda_{345}-\mu_{12})c_\beta s_\beta &
-\frac12{\cal I}m\lambda_5 s_\beta\\
(\lambda_{345}-\mu_{12})c_\beta s_\beta & \lambda_2 s_\beta^2+\mu_{12}c_\beta^2 &
-\frac12{\cal I}m\lambda_5 c_\beta\\
-\frac12{\cal I}m\lambda_5 s_\beta & -\frac12{\cal I}m\lambda_5 c_\beta & -{\cal R}e\lambda_5 +\mu_{12}
\end{matrix}\right).
\ee
We define the diagonalizing matrix $R$:
\be
R{\cal M^2}R^T={\rm diag}(m_{h_1}^2,m_{h_2}^2,m_{h_3}^2),
\ee
so that
\be
\left(\begin{matrix} h_1 \\ h_2\\ h_3\end{matrix}\right)=
R\left(\begin{matrix} H_1^0 \\ H_2^0\\ A^0\end{matrix}\right).
\ee
For fields with no diagonal coupling to $\Phi_2$, such as $Y^2_{ee}=0$, we have
\be
\kappa_e=R_{h1}/c_\beta,\ \ \ \tilde\kappa_e=-R_{h3}t_\beta.
\ee
For fields with no diagonal couplings to $\Phi_1$, $Y^1_{ff}=0$, we have
\be
\kappa_f=R_{h2}/s_\beta,\ \ \ \tilde\kappa_f=-R_{h3}\cot\beta.
\ee
We learn that for the electron
\be
\tilde\kappa_e/\kappa_e=(R_{h3}/R_{h1})\sin\beta.
\ee
In particular, an upper bound on $\tilde\kappa_e/\kappa_e$ translates into an upper bound on $R_{h3}/R_{h1}$.

In the large $\tan\beta$ limit, and assuming that the diagonal terms in ${\cal M}^2$ are not quasi-degenerate, we obtain
\begin{align}
R_{21}\sim&\frac{1}{t_\beta}\frac{\mu_{12}-\lambda_{345}}{\mu_{12}-\lambda_2},\nonumber\\
R_{23}\sim&\frac{1}{2t_\beta}\frac{{\cal I}m\lambda_5}{\mu_{12}-\lambda_2-{\cal R}e\lambda_5},\nonumber\\
R_{13}\sim&\frac12\frac{{\cal I}m\lambda_5}{{\cal R}e\lambda_5}.
\end{align}
Identifying $h$ with $h_2$, we get
\begin{align}
\kappa_e\sim&\frac{\mu_{12}-\lambda_{345}}{\mu_{12}-\lambda_2},\nonumber\\
\tilde\kappa_e\sim&\frac{(1/2){\cal I}m\lambda_5}{\mu_{12}-\lambda_2-{\cal R}e\lambda_5},\nonumber\\
\frac{\tilde\kappa_e}{\kappa_e}=&{\cal O}\left(\frac{{\cal I}m\lambda_5}{\mu_{12}-\lambda_{345}}\right).
\end{align}

One can also express the results in terms of the two rephasing invariant complex phases:
\begin{align}
\delta_1=&{\rm arg}\left[\lambda_5^*(m_{12}^2)^2\right],\\
\delta_2=&{\rm arg}\left[\lambda_5^*(m_{12}^2)v_1v_2^*\right].\nonumber
\end{align}
The minimum equations relate these two phases, so that only one is independent. For $|m_{12}^2|t_\beta\gg v^2$ and small phases, we have $\delta_2\simeq\delta_1$, and
\be
\frac{\tilde\kappa_e}{\kappa_e}\simeq\frac{|\lambda_5|\sin\delta_1}{|m_{12}^2/v^2|t_\beta}.
\ee

We conclude that the phases in the scalar potential, which can a-priori be ${\cal O}(1)$, must be smaller -- in the case that $\kappa_e={\cal O}(10^2)$ -- than ${\cal O}(10^{-4})$. Let us note that even if $\kappa_e={\cal O}(1)$, the scalar potential of a 2HDM should be CP conserving to the level of $10^{-2}$.

\subsection{The Yukawa sector}
In this subsection we assume that CP is a good symmetry of the scalar potential, such that the neutral mass eigenstates are the even ($h$ and $H$) and odd ($A$) CP eigenstates. We ask whether strong enhancement of $\kappa_e^X$ makes the bounds from one-loop contributions to $d_e$ competitive with the bounds from two-loop contributions (\ref{eq:keexp}).

The one-loop flavor-conserving Higgs contribution to the electron EDM is given by (for example, see~\cite{Broggio:2014mna})
\be\label{eq:EDM}
d_e = -\frac{e y_e^2 \tilde{\kappa}_e \kappa_e}{(4\pi)^2}\frac{m_e}{m_h^2}\left(\ln\frac{m_e^2}{m_h^2}+\frac{7}{6}\right),
\ee
Imposing the upper bound on $d_e$ of Eq. (\ref{eq:deexp})~\cite{Baron:2013eja}, we arrive at
\be\label{eq:boundkappa}
\kappa_e\tilde{\kappa}_e \lsim 1.1\times 10^5.
\ee
Since $\kappa_e\tilde{\kappa}_e \leq \frac{1}{2}(|\kappa_e|^2+|\tilde\kappa_e|^2)$, Eq.~(\ref{eq:boundkappa}) is automatically satisfied when the bound on $\mu_{ee}$ of Eq. (\ref{eq:muee})~\cite{Khachatryan:2014aep} is imposed.

As for the contribution of the heavy scalar loops, we have
\be
d_e^H &=& -\frac{e y_e^2 \tilde\kappa^H_e \kappa^H_e}{(4\pi)^2}\frac{m_e}{m_H^2}\left(\ln\frac{m_e^2}{m_H^2}+\frac{7}{6}\right); \\ \nonumber
d_e^A &=& \frac{e y_e^2 \tilde\kappa^A_e \kappa^A_e}{(4\pi)^2}\frac{m_e}{m_A^2}\left(\ln\frac{m_e^2}{m_A^2}+\frac{11}{6}\right); \\ \nonumber
d_e^{H^\pm} &=& -\frac{e y_e^2 \tilde\kappa^{A}_e \kappa^{A}_e}{6(4\pi)^2}\frac{m_e}{m_{H^\pm}^2}.
\ee
We obtain the following bounds:
\be\label{eq:boundkappaH}
\tilde\kappa^{H,A}_e \kappa^{H,A}_e \lsim 7\times 10^6 \left(\frac{m_{H,A}}{1 {\rm TeV}}\right)^2.
\ee
In the scenario where the scalar potential is real and CP violation comes from phases in the Yukawa entries, we found that $\kappa_e^{H,A}$ and $\tilde\kappa_e^{H,A}$ are $t_{\beta-\alpha}$ enhanced compared to $\kappa_e$ and $\tilde \kappa_e$. Thus, for $c_{\alpha-\beta}\ll1$, Eq.~(\ref{eq:boundkappaH}) provides stronger bounds than Eq.~(\ref{eq:boundkappa}). However, this constraint competes with the bound coming from the Barr-Zee diagrams \cite{Barr:1990vd} only if $0<|c_{\beta-\alpha}|\lsim 4\times 10^{-5} \left(
{\rm TeV}/m_S\right)$.

If the heavy scalars are quasi-degenerate, $m_H\approx m_A\approx m_{H^\pm}$, and $t_{\beta-\alpha}\gg 1$ so that $\kappa_e^{H^\pm} = \kappa_e^A\approx \kappa_e^{H} \approx t_{\beta-\alpha}k_e$, the total contribution of scalars at one loop is given by
\be
d_e\approx -\frac{ey_e^2 \kappa_e\tilde\kappa_e}{(4\pi)^2}\frac{m_e}{m_h^2}\left(\ln\frac{m_e^2}{m_h^2}+\frac{7}{6}-
\frac{t_{\beta-\alpha}^2}{2}\frac{m_h^2}{m_H^2}\right).
\ee
%

\section{LHC phenomenology}
\label{sec:lhc}
In Section~\ref{sec:spectrum} we argued that in most of the parameter space relevant to $\kappa_e\gg1$, all scalars should be at the electroweak scale. An exception arises if  $\lambda_7={\cal O}(1)$, but even in this case a large portion of the parameter space is within the reach of the LHC. Hence, we can probe the $\kappa_e\gg1$ scenario in 2HDM indirectly via LHC searches for new scalars.

\subsection{ $A^0,H^0$ and $H^{\pm} $ decay modes}
The Yukawa coupling of $A$ to electrons (\ref{eq:keA}) is enhanced. In order to establish if the $A\to e^+e^-$ decay has a phenomenological impact we need to compare  Eq. (\ref{eq:keA}) with the coupling to other SM fermions:
\be
\frac{y^A_e}{y^A_f}= \frac{y_e}{y_f} \frac{\kappa_e^A} {\kappa_f^A}.
\ee

Our first observation in this regard is that, for $\kappa_e={\cal O}(500)$, $A\to e^+e^-$ will dominate over $A\to f\bar f$ for any fermion $f$ for which $y_1^f=0$. For such fermions,
\be
\frac{y^A_e}{y^A_f}= \frac{y_e}{y_f} t_{\beta_e}^2
\ee
The strongest hierarchy of the SM Yukawa couplings is for $y_e/y_t$. If $y_1^t=0$, then
\be
t_{\beta_e} \gsim 500 \implies y^A_e \gsim y^A_t.
\ee
This condition is met when $\kappa_e\gsim500\sqrt{1-\kappa_V^2}$ and, in particular, for $\kappa_e\gsim500$, as we assume. Obviously, if $y_1^t=0$ implies $y^A_e>y^A_t$, then $y_1^f=0$ guarantees $y^A_e>y^A_f$ for any fermion $f$. Similar conclusions hold for $H^+\to e^+\nu$ and $H^0\to f\bar f$.

Our second observation is that, the ratio of $\Gamma(A\to e^+e^-)/\Gamma(A\to f\bar f)=(y_e/y_f)^2$ for any fermion $f$ for which $y_2^f=0$. Thus, if any second generation fermion (or, obviously, third) has $y_2^f=0$, $A\to e^+e^-$ will have little phenomenological impact. Similar conclusions hold for $H^+\to e^+\nu$ and $H^0\to e^+e^-$. If the $u$-quark and/or the $d$-quark have $y_2^f=0$, then the dielectron decay rate of the heavy scalars will be subdominant to the dijet rate, but not negligible.

Our third observation makes use of Eq. (\ref{eq:higgstomass}), which gives
\be
\frac{\kappa_e^A}{\kappa_f^A}=\frac{\kappa_e-s_{\beta-\alpha}}{\kappa_f-s_{\beta-\alpha}}.
\ee
Thus, if experiments put an upper bound on $\kappa_f$, $\kappa_f^{\rm max}$, then for large $\kappa_e\gg \kappa_f^{\rm max}$, we have
\be
\frac{\kappa_e^A}{\kappa_f^A}\gsim\frac{\kappa_e}{\kappa_f^{\rm max}}.
\ee
Such upper bounds apply already to $f=t,b,\tau,c,\mu$. They prove that, if $\kappa_e={\cal O}(500)$ then $y_2^f\neq0$ for all of these fermions. For the muon case, the present bound is sufficient to guarantee that if $\kappa_e={\cal O}(500)$ then $A\to e^+e^-$ dominates over $A\to \mu^+\mu^-$.

\subsection{Multi-electron signatures}
In the previous subsection we obtained conditions under which the $A,H\to e^+e^-$ and $H^+\to e^+\nu$ are the dominant decay modes of the heavy scalars. Specifically, it is required that the heavy scalar couplings to the third generation fermions are strongly suppressed (which is the case for $y_1^f=0$). The conditions for suppressing the heavy scalar decays into third generation fermions also entail strong suppression of single heavy scalar production, e.g. $gg\to A$ and $gb\to tH^-$.  Furthermore, in the $ \kappa_V\to1$ limit, also the production via vector boson fusion is subdominant.

We distinguish two scenarios with large $\kappa_e$: one where $y_2^e=y_2^d=y_2^u=0$ and consequently also $ \kappa_u, \kappa_d$ are large, which we discuss in the next subsection, and one where only $y_2^e=0$, which we focus on here. In this case, the relevant production modes of heavy scalars  are pair production via the $s$-channel mediation of a virtual electroweak vector boson:
\be
p p \rightarrow \gamma^* \rightarrow H^+ H^- , \; \;  p p \rightarrow W^* \rightarrow H^+ A^0(H^0) ,\; \; p p \rightarrow Z^* \rightarrow A^0H^0.
\ee
Since we assume here that the branching ratio into electrons is dominant, the relevant topologies are the following:
\begin{itemize}
\item Two electrons plus missing energy:
\be
p p \rightarrow \gamma^*/Z^* \rightarrow H^+ H^- \rightarrow e^+ e^- \nu \nu.
\ee
\item Three electron plus missing energy:
\be
p p \rightarrow W^* \rightarrow H^0 (A^0)  H^{\pm} \rightarrow e^+ e^- e^\pm \nu.
\ee
\item Four electrons:
\be
p p \rightarrow Z^* \rightarrow H^0 A^0 \rightarrow e^+ e^- e^+ e^-
\ee
\end{itemize}
Hence multi-electron signatures are the distinctive feature of this scenario.

We are not aware of relevant targeted searches for this topology in 2HDM. However, multi-lepton searches, typically aimed for models of neutrino masses, have been carried out and can be recasted for our topology. In particular, the 13 TeV CMS search \cite{Sirunyan:2017qkz} is a multipurpose analysis which can be  applied to other topologies other than the one originally designed for. A recast was made in Ref.~\cite{Abe:2017jqo} assuming dominance of final multi-muon states and $s_{\beta-\alpha}=1$, and a bound $m_A>640$ GeV was extracted.\footnote{For $ 0.9 <s_{\beta -\alpha} <1 $ the bound does not change significantly, hence we can consider this as the reference value.} Since we do not expect the efficiency to change significantly, we take this bound to be a rough indication of the bound that applies for the multi-electron case. We conclude:
\begin{itemize}
\item The scenario with $ \lambda_{6,7} =0$, where $\kappa_e\gg1$ requires $m_H\simeq m_h$ [see Eq. (\ref{eq:keg67})], is excluded;
\item The scenario with hard breaking of the $Z_2$ symmetry is strongly constrained unless $\lambda_7$ is very large or $\kappa_V$ very close to 1.
\end{itemize}

\subsection{Large first generation Yukawa couplings}
An interesting case arises if one of the Higgs doublets does not couple to the first generation fermions, $y_2^e=y_2^d=y_2^u=0$ and, consequently, $\kappa_e=\kappa_d=\kappa_u$. In this case, the branching ratio of the heavy scalars to dielectrons is considerably smaller than the branching ratio into dijets, but the production cross-section is enhanced via the $s$-channel $q\bar q^\prime\to A,H,H^\pm$. Therefore, dielectron resonance searches can become relevant. (For an interesting proposal of how to probe enhanced $\kappa_{u,d}$, see Ref. \cite{Yu:2016rvv}.)

In order to explore the phenomenological implications of such a framework, we further assume that, to a good approximation, the other Higgs doublet does not couple to the second and third generations. Thus, we consider the following scenario:
\begin{align}\label{eq:k1k23}
\kappa_e&=\kappa_u=\kappa_d,\nonumber\\
\kappa_t&=\kappa_b=\kappa_\tau=\kappa_c=\kappa_s=\kappa_\mu,
\end{align}
and
\be\label{eq:kvkekt}
\kappa_V=\frac{1+\kappa_e\kappa_t}{\kappa_e+\kappa_t}.
\ee
Therefore, the couplings of the extra scalars to the whole first generation are $t_{\beta}$ enhanced, so this is the relevant parameter to evaluate the constraints, together with the mass $m_H$.

The mass window $120-150$ GeV is probed by the CMS search for  $h\to e^+e^-$  \cite{Khachatryan:2014aep}  and is excluded for $ t_{\beta} >800-900$.  For $ \kappa_V \sim 1$ this implies that also moderate values for $\kappa_{e,u,d}$ are ruled out.

Heavier masses are constrained by  dilepton searches from both ATLAS \cite{Aad:2014cka} and CMS \cite{Khachatryan:2014fba}. These searches look for both dielectron and dimuon final states, targeting as benchmark models new $Z'$ gauge bosons. Their limit can be straightforwardly applied to our scenario, as long as they are presented separately for dimuons and dielectrons. We obtain the exclusion limits on our scenario from the 8 TeV data. The current 13 TeV data do not change the picture significantly. CMS published so far results with only $2.9$ fb$^{-1}$ data from run II \cite{Khachatryan:2016zqb}, while  ATLAS \cite{Aaboud:2017buh} present only combined results from dielectron and dimuon channels.

\begin{figure}[t]
 \begin{center}
  \includegraphics[width=0.53\textwidth]{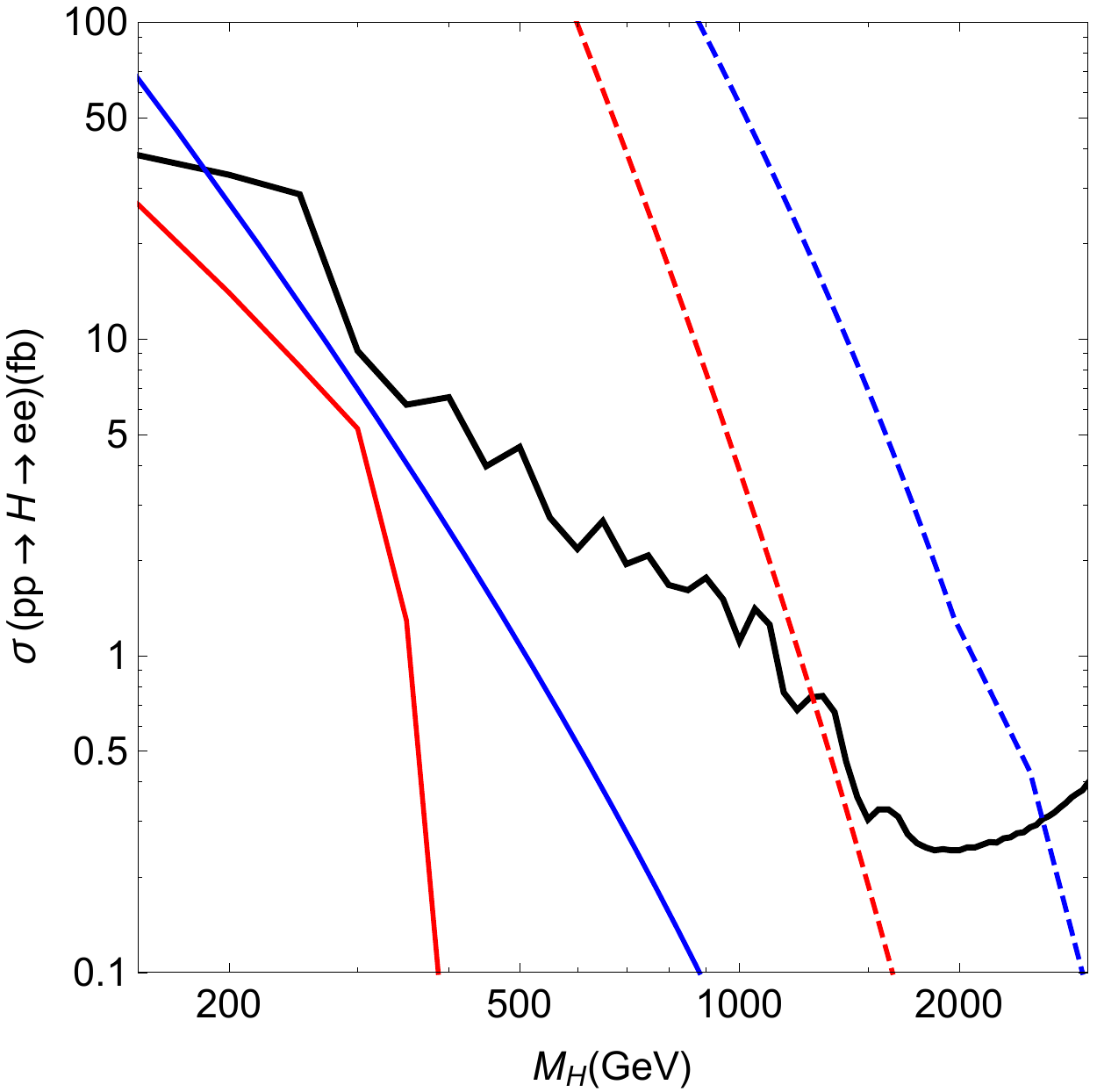} \
 \end{center}
 \caption{$\sigma(pp\to H,A\to e^+e^-)$ as a function of $m_H=m_A$ for $\kappa_e=500$.
 The region above the black curve is excluded by the 8 TeV ATLAS search \cite{Aad:2014cka}. The theoretical predictions corresponding to the scenario defined by Eqs. (\ref{eq:k1k23}) and (\ref{eq:kvkekt}) are given in blue curves, with $t_\beta=1000$ (solid) or $4\times10^4$ (dashed). The theoretical predictions corresponding to $y_t^{A,H}=1$ are given in red curves, with $\tan\beta=1000$ (solid) or $4\times10^5$ (dashed).}
\label{fig:kukd}
\end{figure}

Fig.~\ref{fig:kukd} presents the constraints from resonant dielectron 8 TeV searches in the 150 GeV - 3 TeV range. The experimental exclusion curve, based on \cite{Aad:2014cka}, is given in solid black (similar results follow from the CMS search \cite{Khachatryan:2014fba}). We computed the leading order cross-section for  $q\bar q^\prime\to H,A\to e^+e^-$ using the NNPDF2.3 LO pdf  Mathematica package \cite{Ball:2012cx,Hartland:2012ia}. We use two values for $t_\beta$: First, $t_{\beta}=1000$ (solid blue curve), which is close to the minimum value constrained by this search. Second, $t_\beta=4\times10^4$ (dashed blue curve), which corresponds to $1/y_d$, the maximum allowed value from perturbativity.

We reach the following conclusions:
\begin{itemize}
\item The CMS search \cite{Khachatryan:2014aep} rules out the existence of $H^0$ and $A^0$ with mass in the $120-150$ GeV range and $\tan\beta>900$. For $\kappa_e=500$, this implies in turn $\kappa_V>0.83$.  (For $\kappa_e=50$, this implies $\kappa_V>0.998$).
\item
The ATLAS search \cite{Aad:2014cka} rules out $m_{H,A}<200\ (2500)$ GeV for $\tan\beta>10^3\ (1/y_d)$.
\end{itemize}
Hence, the scenario with $\lambda_{6,7}=0$ is almost ruled out for all values of $t_{\beta}$.
We expect in the future the limit in the high $t_{\beta} $ to become more stringent pushing towards higher values of $ \lambda_7$ or $\kappa_V$ closer to one.

\subsection{Production via gluon gluon fusion} \label{gluonfusion}
Another scenario which gives rise to resonant production via gluon gluon fusion of the extra scalars and subsequent decay into $e^{+} e^{-}$ corresponds to  $ y_1^{t,b} \neq 0$. As a case study we consider the following benchmark point:
\be\label{eq:kAt}
\kappa^{A,H}_t=\kappa^{A,H}_b=\kappa^{A,H}_\tau=1,\ \ \
%
\kappa^{A,H}_e &=  t_{\beta}.
\ee
As concerns $A$ and $H$ production, only $\kappa^{A,H}_t$ is relevant. As concerns the decay, especially for $m_{H,A}<2m_t$, also $\kappa^{A,H}_{\tau,b}$ play a role. We consider then the same LHC searches described in the previous subsection. We compute the cross section also with  the NNPDF2.3 LO pdf  Mathematica package \cite{Ball:2012cx,Hartland:2012ia}. The results are presented in Fig.~\ref{fig:kukd} in red curves for $t_\beta=1000$ (solid) and $t_\beta=1/y_e$ (dashed). We reach the following conclusions:
\begin{itemize}
\item The searches do not constrain this scenario in the ``low" $\tan\beta$ region ($t_{\beta}<1000$) in both the low mass \cite{Khachatryan:2014aep} and the high mass \cite{Aad:2014cka} ranges.
(We do expect mild constraints from present searches when NLO corrections are incorporated.)
\item The large $\tan\beta$ region, $t_{\beta}\sim 1/y_e$, is excluded up to $m_{H,A}\sim1.8$ TeV.
\end{itemize}
This scenario is then less constrained compared to the previous one.
The reason is that the production cross section does not increase with $t_{\beta}$, but only the branching ratio which is, however, very suppressed when the decay mode into two tops is open.

\section{2HDM in the literature}
\label{sec:models}
Some of our results have implications that are generic to 2HDM, and go beyond the specific scenario of $\kappa_e\gg1$ or $\kappa_e\ll1$. We here give a brief survey of models that have been proposed in the literature, and their relation with our findings.

\subsection{Separating the third generation from the first two}
In the 2HDM of Ref. \cite{Ghosh:2015gpa}, one of the Higgs doublets does not couple to the third generation quarks, while the other has negligibly small diagonal couplings to the first two generations. (See Ref. \cite{Altmannshofer:2016zrn} for a related scenario.) Thus, in this model,
\begin{align}
\kappa_l&\equiv\kappa_u=\kappa_d=\kappa_s=\kappa_c,\nonumber\\
\kappa_h&\equiv\kappa_b=\kappa_t,
\end{align}
and
\be
\kappa_V=\frac{1+\kappa_l\kappa_h}{\kappa_l+\kappa_h}.
\ee
The authors aim to have $\kappa_l\ll1$. Given that it is experimentally known that both $\kappa_V$ and $\kappa_h$ are ${\cal O}(1)$ then, in this case, the model predicts $\kappa_V\kappa_h\simeq1$.

The model further has $\lambda_6=\lambda_7=0$ and is thus subject to the analysis of Section \ref{sec:l6l7}. The model however further assumes $\lambda_3=\lambda_4=\lambda_5=0$. The requirement for $\kappa_e\gg1$ is still (\ref{eq:keg67}), but the requirement for $\kappa_e\ll1$ is no longer (\ref{eq:kes67}). Examining Eq. (\ref{eq:spectrum67}), we learn that $\kappa_e\ll1$ implies $\kappa_e/\kappa_V\simeq m_A^2/(m_H^2-m_h^2)$, so that $m_A^2\ll m_H^2-m_h^2$ is required. Since in this case
\be
m_H^2+m_h^2=m_A^2+v^2(\lambda_1 c_\beta^2+\lambda_2 s_\beta^2),
\ee
we must have $m_H^2={\cal O}(v^2)$ and $m_A^2\ll v^2(\lambda_1 c_\beta^2+\lambda_2 s_\beta^2)$. We learn that the scalar spectrum is light. A problem might arise however given that for $\lambda_4=\lambda_5=0$ we have $m_{H^\pm}^2=m_A^2$, and there is a rather strong lower bound on $m_{H^\pm}^2$.

In the 2HDM of Ref. \cite{Altmannshofer:2015esa}, one of the Higgs doublets, $\phi$, does not couple to the first two generations, while the other, $\phi^\prime$, has small couplings to the third generation. Thus, in this model, $y_\phi^e=y_\phi^\mu=0$, and the model should have $\kappa_e=\kappa_\mu$.
Various equations of Ref. \cite{Altmannshofer:2015esa} can be evaluated to get Eq. (\ref{eq:classtwo}) as an approximate relation.

In the 2HDM of Ref. \cite{Botella:2016krk}, the masses of the first two quark generations come from dimension-six terms. Thus, the model predicts
\be
\kappa_c=\kappa_s=\kappa_d=\kappa_u=3.
\ee
%

\subsection{Separating the muon from the other fermions}
In the $\mu$2HDM of Ref. \cite{Abe:2017jqo}, one of the Higgs doublets, $H_2$, couples to the up sector, to the down sector, and to $e$ and $\tau$. The other Higgs doublet, $H_1$, couples to only $\mu$. Thus, for the quarks, this is an NFC model, with the well known consequences of that. In the charged lepton sector, however, we have the situation where
\be
y_2^\mu=0,\ \ y_1^e=y_1^\tau=0.
\ee
From the discussion in Section \ref{sec:kappaf}, the following relations hold:
\be
\kappa_\tau=\kappa_e,\ \ \ \kappa_V=\frac{1+\kappa_\mu\kappa_\tau}{\kappa_\mu+\kappa_\tau}.
\ee
Thus, the experimental information that $\kappa_V$ and $\kappa_\tau$ are close to $1$, implies that so is not only $\kappa_e$ but also $\kappa_\mu$.

\section{Conclusions}
\label{sec:conclusions}
We studied the implications of a strongly enhanced Higgs-electron Yukawa coupling, such that $h\to e^+e^-$ will be within reach of ATLAS/CMS in the near future, $\kappa_e\equiv Y_{ee}/Y_{ee}^{\rm SM}={\cal O}(500)$. We focussed on two Higgs doublet models (2HDM). Given the experimental measurements of $h\to\tau^+\tau^-$ and the upper bound on $h\to \mu^+\mu^-$, such a strong enhancement of $Y_{ee}$ excludes also 2HDM with natural flavor conservation (NFC). We thus explored generic 2HDM.

We suggested a basis for the two scalar doublets which is particularly convenient to study implications of enhanced electron Yukawa coupling. Our proposed basis can be straightforwardly generalized to any other fermion with a Yukawa coupling that is very different from the SM prediction.

Our main findings are the following:
\begin{itemize}
\item Case I: For two fermions with vanishing Yukawa couplings to one and the same of the two Higgs doublets, the enhancement factors are the same, $\kappa_{f_1}=\kappa_{f_2}$. Furthermore, the modification factors of their couplings to the heavy scalars, $H,A,H^\pm$, are the same.
\item Case II: For two fermions with vanishing Yukawa couplings to two different Higgs doublets, the enhancement factors fulfill a relation, $\kappa_V=(1+\kappa_{f_1}\kappa_{f_2})/(\kappa_{f_1}+\kappa_{f_2})$. Similarly, their couplings to heavy scalars fulfill predictive relations.
\item If the Yukawa coupling to the Higgs is enhanced, $Y^h_{ff}\gg\sqrt2 m_f/v$, the Yukawa coupling to the heavy scalars is even more strongly enhanced (by order $\tan(\alpha-\beta)$).
\item In case II, if $\kappa_{f_1}\gg1$ so that the Yukawa couplings of $f_1$ to $A,H,H^\pm$ are very large, $\kappa_{f_2}\approx1$ while the couplings of $f_2$ to the heavy scalars are suppressed.
\item In models with only soft breaking of a $Z_2$ symmetry in the scalar potential, a large deviation of $\kappa_e$ from 1 requires a light scalar spectrum. With hard breaking, there is an interesting range where $v^2\ll m_A^2\ll v^2\kappa_e/\sqrt{1-\kappa_V^2}$ where such deviation is still possible.
\item For $\kappa_e\gsim1$ CP symmetry should hold to a good approximation [${\cal O}(10^{-2}/\kappa_e)$] in both the scalar potential and the Yukawa couplings.
\item Large regions of the parameter space of 2HDM models with $\kappa_e\gg1$ are probed by ATLAS/CMS searches for deviations of the $e^+e^-$ mass spectra from the SM.
\item Searches for $e^+e^-e^+e^-$ and $e^+e^-e^\pm\not E_T$ topologies will provide sensitive probes of this scenario.
\end{itemize}

\acknowledgments
We thank Jo\~ao Silva and Ryosuke Sato for useful discussions.
YN is the Amos de-Shalit chair of theoretical physics.
YN is supported by grants from the Israel Science Foundation (grant number 394/16), the United States-Israel Binational Science Foundation (BSF), Jerusalem, Israel (grant number 2014230), the I-CORE program of the Planning and Budgeting Committee and the Israel Science Foundation (grant number 1937/12), and the Minerva Foundation.


\end{document}